\documentstyle[12pt]{article}

\fnsymbol{footnote}

\newcommand{\be}{\begin{equation}}
\newcommand{\ee}{\end{equation}}
\newcommand{\bea}{\begin{eqnarray}}
\newcommand{\eea}{\end{eqnarray}}

\begin{document}

\thispagestyle{empty}
\setcounter{page}{0}

\begin{center} 
\Large \bf
The non-dynamical r-matrices of the degenerate Calogero-Moser models\\
\end{center}

\vspace{.1in}

\begin{center}
L. Feh\'er and B.G. Pusztai\\
\vspace{0.3in}
{\em Department of Theoretical Physics, University of Szeged\\
Tisza Lajos krt 84-86, H-6720 Szeged, Hungary \\
e-mail: lfeher@sol.cc.u-szeged.hu }
\end{center}

\vspace{.3in}

\begin{center} {\bf Abstract} \end{center}

A complete description of the non-dynamical 
r-matrices of the degenerate Calogero-Moser models 
based on $gl_n$ is presented.
First the most general momentum independent r-matrices
are given for the standard Lax representation of these systems 
and  those r-matrices whose coordinate dependence can be 
gauged away are selected. 
Then the constant r-matrices  
resulting from gauge transformation are determined  and are 
related to well-known  r-matrices.
In the hyperbolic/trigonometric case a
non-dynamical r-matrix equivalent 
to a real/imaginary multiple of the Cremmer-Gervais 
classical r-matrix is found.
In the rational case the constant r-matrix
corresponds to the antisymmetric solution of the 
classical Yang-Baxter equation associated with  
the Frobenius subalgebra of $gl_n$ consisting of   
the matrices with vanishing last row.
These claims are consistent with previous results 
of Hasegawa and others, which imply  that 
Belavin's elliptic r-matrix and its degenerations appear 
in the Calogero-Moser models.
The advantages of our analysis are that it is elementary
and also clarifies  
the extent to which the constant r-matrix is unique  
in the degenerate cases.

\newpage

\section{Introduction}

The purpose of this paper is to provide a complete 
description of 
the non-dynamical, constant r-matrices of the standard Calogero-Moser models 
\cite{C,M} 
associated  with degenerate potential functions,
which can be obtained by gauge transformations of their usual Lax 
representation.
A preliminary account of a part of this work is contained in \cite{FP}.

The Calogero-Moser type many particle systems (for a review, see \cite{P}) 
have been much studied recently due to their fascinating 
mathematics and applications ranging from 
solid state physics to Seiberg-Witten theory.
The definition of these models involves
a root system and a potential function depending on the 
inter-particle `distance'.
The potential is given either by the Weierstrass ${\cal P}$-function or 
one of its (hyperbolic, trigonometric or rational) degenerations.  
The classical equations of motion of the  models admit 
Lax representations,
\be
\dot{L}=[L,M], 
\label{1}\ee 
which underlie their integrability.
A  Lax representation  of the Calogero-Moser  
models based on the root systems of the classical Lie algebras 
was found by Olshanetsky and Perelomov \cite{OP}
using symmetric spaces. 
Recently new Lax representations for these systems 
as well as their exceptional Lie algebraic 
analogues and twisted versions have been constructed \cite{dHP,BCS}.

In general \cite{BV}, Liouville ingtegrability can be understood as
a consequence   of the Poisson brackets of the Lax matrix 
having  the r-matrix form,
\be
\{ L_1, L_2 \} = \{ L^\mu, L^\nu\} T_\mu \otimes T_\nu = 
[r_{12}, L_1] - [r_{21}, L_2],
\label{2}\ee
where $r_{12}=r^{\mu \nu} T_\mu \otimes T_\nu$ 
($r_{21}=r^{\mu \nu } T_\nu \otimes T_\mu$) 
with some 
constant matrices $T_\mu$.
The components $L^\mu$ of the Lax matrix 
$L=L^\mu T_\mu$ encode the phase space variables,
and the components $r^{\mu\nu}$ of the classical r-matrix 
may in general depend on the same variables as $L_1=L\otimes 1$ 
and $L_2= 1 \otimes L$.
Of course, $L$ and $r$ depend also on a spectral parameter
in general, but this does not occur for
the systems of our interest, and thus is suppressed in (2).
When the r-matrix really does depend on the phase space variables,
one says that it is `dynamical'.

The classical r-matrix has been calculated 
first for the standard Lax representation of the $gl_n$
Calogero-Moser systems 
associated with degenerate potentials \cite{AT}, 
and then for  
Krichever's \cite{Kri} spectral parameter 
 dependent Lax matrix in the elliptic case \cite{Skly,BS}.
The r-matrices found in these papers are dynamical, but depend only on
the coordinates of the particles.
These r-matrices  have been re-derived by means of Hamiltonian 
reduction in \cite{ABT,AM}, 
and in a very recent paper \cite{FW} they have been generalized 
explicitly for the $BC_n$ system as well as for all classical Lie algebras.
In the physically most interesting $gl_n$ case, dynamical r-matrices
have also been found \cite{AR,NKSR,S} for the relativistic deformations of 
the Calogero-Moser models introduced by Ruijsenaars and Schneider \cite{RS}.
Then the quantization of the non-relativistic  \cite{ABB}
and the relativistic models \cite{FeWa,AF,ACF} 
has  been investigated in a new 
framework based on quantum dynamical R-matrices.

The above developments have close connections 
with the new theory of dynamical r-matrices and 
associated quantized structures  reviewed in \cite{ES}.
However, since the present understanding  of most integrable  systems
involves constant (i.e.~`non-dynamical') r-matrices, which form a direct link to  
Poisson-Lie groups and quantum groups \cite{CP},
it is natural to ask if the Lax representation 
of the Calogero-Moser models can be chosen in such a way to exhibit
non-dynamical  r-matrices.
The obvious way to search for new Lax representations with this property
is to perform gauge transformations on the usual Lax representations.
In the elliptic case of the standard $gl_n$ models a new Lax 
representation associated with Belavin's \cite{Bel} constant elliptic
r-matrix  has recently been found in this way \cite{Hou}. 
To be more precise, the results of \cite{Hou} are already contained
in a somewhat less explicit form in the seminal paper by Hasegawa \cite{H},
where the commuting Ruijsenaars operators \cite{R}  have been interpreted
as commuting transfer matrices based on a realization of the
$RLL =LLR$  
relation with Belavin's elliptic R-matrix 
and certain difference $L$-operators. 
In fact, the dynamical twisting and the 
classical and non-relativistic limits of the
$L$-operator leading to Krichever's Lax matrix for the elliptic
Calogero-Moser model are indicated in \cite{H} (see also \cite{ACF}).
Then in the paper \cite{AHZ}
some delicate limit procedures have been considered, whereby 
non-dynamical R-matrices can be obtained for the trigonometric
degenerations of the Ruijsenaars-Schneider and Calogero-Moser models.
The resulting R-matrix was found to be 
non-unique, one possibility \cite{AHZ} being the spectral parameter 
independent Cremmer-Gervais R-matrix discovered in a
different context in \cite{CG}. 

It is clear from the above that Lax representations for
the degenerate Calogero-Moser models with non-dynamical
r-matrices can be obtained by taking limits of
Hasegawa's $RLL=LLR$ relation. However, the details
of the admissible limiting procedures appear rather complicated 
and  the starting point requires familiarity with quite advanced 
results.
In this circumstance, it might be worthwhile to understand 
the possible 
non-dynamical r-matrices 
also from an elementary viewpoint.
This is the objective of the present paper,
where we aim to perform a self-contained, systematic analysis
of the gauge transformations of the usual Lax representation
of the degenerate Calogero-Moser models that lead to constant r-matrices.

The organization and the main results of our work
 are as follows.
First, we describe the most general momentum independent dynamical r-matrices 
for the standard Lax representation in section 2.
This amounts to a slight but necessary generalization of 
the Avan-Talon \cite{AT} 
r-matrix as given by Theorem 1.
Second, we select those dynamical r-matrices that 
become constant by a gauge transformation (defined by eq.~(\ref{Rprime})) 
and determine the
corresponding  `gauge potentials'$A_k(q)$.
This is the content of section 3, in particular Proposition 2 and Theorem 3.
Third, in section 4 we compute explicitly the gauge transformations $g(q)$
(from eq.~(\ref{defA})) 
and the resulting most general constant r-matrix, which 
is given by Theorem 6.
It turns out that in the rational case the constant  r-matrix 
is conjugate to the antisymmetric solution of the 
classical Yang-Baxter equation  
that belongs to the Frobenius subalgebra of $gl_n$ consisting of   
the matrices with vanishing last row \cite{BD}.
In the hyperbolic/trigonometric cases the $sl_n$-part of 
the most general $gl_n\wedge gl_n$-valued constant r-matrix 
(see Proposition 7) 
is equivalent 
to a multiple of the Cremmer-Gervais classical r-matrix \cite{CG,GG},
and it can also be   made equal to it by a choice of the gauge transformation.
This identification of the constant Calogero-Moser r-matrices
is presented in section 5.
The main results are summarized once more in the conclusion,
which occupies section 6.
Except for the notations introduced in section 2,
this final section is self-contained and it may be useful 
to consult it before reading the main text.  
The details of some proofs are contained in three appendices.

The outcome of our direct analysis of the degenerate 
Calogero-Moser models is consistent 
with the previous results \cite{H,AHZ,Hou}.
In addition to the advantage 
that our analysis is elementary,
we also clarify the extent to which the constant r-matrix 
is unique  in the degenerate cases.
In principle, this uniqueness question 
cannot be answered by studying the limits of the elliptic case,
even though in the final analysis it follows  
that all our constant r-matrices can be 
regarded as various degenerations (see also \cite{EH}) of Belavin's 
elliptic r-matrix.
  
\section{Momentum independent dynamical r-matrices}

The standard (degenerate) Calogero-Moser-Sutherland models 
are defined by the Hamiltonian
\be
h=\frac{1}{2}\sum_{k=1}^n p_k^2 +  \sum_{k<l} v(q_k-q_l),
\label{3}\ee
where $v$ is given as
\be
v(x)=\left\{
\begin{array}{cc} 
x^{-2}, &\mbox{rational case}\\
a^2 \sinh^{-2}(ax),&\mbox{hyperbolic case}\\
a^2 \sin^{-2}(ax),&\mbox{trigonometric case.}\end{array}
\right. 
\label{4}\ee
One has the canonical Poisson brackets
$\{ p_k, q_l\} =\delta_{k,l}$, the coordinates 
are restricted to a domain in ${\mathbf R}^n$ where $v(q_k-q_l)<\infty$,  
and $a>0$ is a parameter.

Let us fix the following notation for  elements of the Lie algebra $gl_n$:
\be
H_k:= e_{kk},\,\,
E_{\alpha}:= e_{kl},\,\,
H_\alpha:=(e_{kk}-e_{ll}),\,\,
K_\alpha:= (e_{kk} + e_{ll})
\,\,\hbox{for}\,\,
\alpha =\lambda_k -\lambda_l\in \Phi.
\label{glnconv}\ee
Here $\Phi=\{ (\lambda_k-\lambda_l) \vert k\neq l\}$ is the set of 
roots of $gl_n$, $\lambda_k$ operates on a  diagonal matrix,
 $H={\mathrm{diag}}(H_{1,1},\ldots,H_{n,n})$ as 
$\lambda_k(H)=H_{k,k}$, and $e_{kl}$ is the $n\times n$ elementary 
matrix whose $kl$-entry is $1$. 
Moreover, we denote the standard Cartan subalgebra of 
$sl_n\subset gl_n$ as ${\cal H}_n$,
and put $p=\sum_{k=1}^n p_k H_k$, $q=\sum_{k=1}^n q_k H_k$,
${\bf 1}_n=\sum_{k=1}^n H_k$. 

{}From the list of known Lax representations 
we  consider the original one \cite{C,M} for which $L$ is the $gl_n$ valued 
function 
\be
L(q,p)=p+ \sqrt{-1}\sum_{\alpha\in \Phi} w(\alpha(q)) E_\alpha,
\label{6}\ee
where the real function $w$ is chosen according to  
\be
w(x)=\left\{
\begin{array}{c} 
x^{-1} \\
a \sinh^{-1}(ax)\\
a \sin^{-1}(ax).
\end{array}
\right.
\label{7}\ee
Then the function  
\be
F:= - {w'\over w}
\label{F}\ee
enjoys the important identities
\be
F'= - w^2,
\label{derprop}\ee 
\be
F(x) + F(y)= {w(x) w(y)\over w(x+y) },
\label{lemma2}\ee 
\be
F(x-y)\left( F(x)-F(y)\right) + F(x) F(y)={\cal B},
\label{lemma1}\ee
where, respectively to the cases above, 
\be
{\cal B}=\left\{
\begin{array}{c} 
0 \\
a^2\\
-a^2.
\end{array}
\right.
\label{C}\ee
For any real function $f$ (like $v$, $w$ or $F$),   we  
introduce the functions $f_k$ and $f_\alpha$  of $q$ as
\be
f_k(q):= f(q_k),\qquad
f_\alpha(q)= f(\alpha(q)),
\label{f-sub}\ee
and sometimes write $f_{kl}$ for $f_\alpha$ if  
$\alpha=(\lambda_k-\lambda_l)$.  
As an $n\times n$ matrix $L_{k,l}= p_k \delta_{k,l} 
+\sqrt{-1} (1-\delta_{k,l}) w(q_k-q_l)$,
but $L$ can also be used in any other representation of $gl_n$.
The r-matrix corresponding to this $L$ was studied by 
Avan and Talon \cite{AT},
who found that it is necessarily dynamical, and may be chosen 
so as to depend on the coordinates $q_k$ only.
We next describe a slight generalization of their result.

\smallskip 
\noindent {\bf Theorem 1.} {\em 
The most general $gl_n\otimes gl_n$-valued r-matrix 
that satisfies (\ref{2}) with the Lax matrix in (\ref{6}) 
and depends only on $q$ is given by
\be
r(q)= -\sum_{\alpha\in \Phi} F_\alpha(q) 
E_\alpha\otimes E_{-\alpha}
+{1\over 2} \sum_{\alpha\in \Phi}
w_\alpha(q) (C_\alpha(q) - K_{\alpha}) \otimes E_\alpha
+ {\bf 1}_n \otimes Q(q),
\label{8}\ee
where the $C_{\alpha}(q)$ are 
${\cal H}_n\subset sl_n$ valued functions subject to the conditions
\be
C_{-\alpha}(q)=- C_\alpha(q),
\qquad
\beta( C_\alpha(q)) = \alpha( C_\beta(q))
\quad
\forall \alpha,\beta\in \Phi 
\label{9}\ee
and $Q(q)$ is an arbitrary $gl_n$-valued function.
}
\smallskip

\noindent
{\em Remarks.}
The functions $C_\alpha$ can be given arbitrarily for the simple roots,
and are then uniquely determined for the other roots by (\ref{9}).
The r-matrix found by Avan and Talon \cite{AT}
is recovered from (\ref{8}) with $C_\alpha\equiv 0$;
and we refer to $r(q)$ in (\ref{8}) as 
the {\em Avan-Talon r-matrix in its general form}. 
Given that this holds for the Avan-Talon r-matrix,
the fact that $r(q)$ above satisfies (\ref{2}) with any $Q(q)$ 
and $C_\alpha(q)$ subject to (\ref{9}) is easy to verify.
Theorem  1 can be proved by a careful calculation
along the lines of \cite{BS}.
For the details, see appendix A. 

\section{Is $r(q)$ gauge equivalent to a constant?}

A gauge transformation of a given Lax representation (\ref{1}) 
has the form 
\be
L \mapsto  L'=g L g^{-1},
\qquad
M\mapsto  M' = gM g^{-1} -  {d g\over dt} g^{-1},
\label{10}\ee
where $g$ is an invertible matrix function on the phase space.
If $L$ satisfies (\ref{2}), then $L'$ will have similar Poisson
brackets with a transformed r-matrix $r'$.
The question now is whether one can remove the $q$-dependence of any of the
r-matrices in (\ref{8}) by a gauge transformation.
It is natural to assume this gauge transformation to be $p$-independent,
i.e.~defined by some function $g:q\mapsto g(q)\in GL_n$.
In this case we find that 
\be
\{ L'_1, L'_2 \} = 
[r'_{12}, L'_1] - [r'_{21}, L'_2]
\label{2'}\ee
holds with  
\be
r'(q)= \left( g(q)\otimes g(q)\right)\left( r(q) + 
\sum_{k=1}^n A_k(q) \otimes H_k\right) 
\left( g(q) \otimes g(q)\right)^{-1},
\label{Rprime}\ee
\be
A_k(q):= -g^{-1}(q) \partial_k g(q), 
\qquad 
\partial_k:= {\partial \over \partial q_k}. 
\label{defA}\ee 
The meaning of this formula is that if $r(q)$ is the most general 
$p$-independent r-matrix for which $L$ (\ref{6}) satisfies (\ref{2}),
then $r'(q)$ has the analogous property in relation to $L'$. 
 
We wish to find $r(q)$ and $g(q)$ such that $\partial_k r'=0$.
On account of (\ref{Rprime}) this  is equivalent to
\be
\partial_k (r + \sum_{l=1}^n A_l \otimes H_l) 
+[ r + \sum_{l=1}^n A_l \otimes H_l , 
A_k \otimes {\bf 1}_n + {\bf 1}_n \otimes A_k] =0.
\label{13}\ee
By using (\ref{defA}), whereby    
\be
{\partial}_k A_l - {\partial}_l A_k + [A_l, A_k]=0,
\label{puregauge}\ee 
it is useful to rewrite (\ref{13}) as 
\be
{\partial}_k r + \sum_{l=1}^n {\partial}_l A_k \otimes H_l 
+ [r, A_k \otimes {\bf 1}_n + {\bf 1}_n \otimes A_k ] 
+ \sum_{l=1}^n A_l \otimes [H_l,A_k]=0.
\label{condonA}\ee  
Our strategy is to first find $A_k(q)$ 
and $r(q)$ from eqs.~(\ref{puregauge}),
(\ref{condonA}), and then determine $g(q)$ and 
the resulting constant r-matrix. 
For this we now parametrize $A_k$ as
\be
A_k(q)= \sum_{l=1}^n A_k^l(q) H_l + 
\sum_{\alpha\in \Phi} A_k^\alpha(q) E_\alpha,
\label{paramA}\ee
and expand the r-matrix from Theorem 1 in the form 
\be
r(q)= -\sum_\alpha F_\alpha(q) E_\alpha \otimes E_{-\alpha}
+ \sum_{i,\alpha} r_i^{\alpha}(q) H_i \otimes E_\alpha 
+ \sum_i Q^i(q) {\bf 1}_n \otimes H_i.
\label{rexpand}\ee
We here have 
\be
r_i^\alpha(q) = Q^\alpha(q) + {1\over 2}w_\alpha(q) 
{\mathrm tr}\left(H_i (C_\alpha(q) - K_\alpha)\right),  
\label{ria}\ee     
\be
Q(q)=\sum_{i=1}^n Q^i(q) H_i + 
\sum_{\alpha\in \Phi} Q^\alpha(q) E_\alpha,
\label{Qexpand}
\ee
where $Q(q)$, $C_\alpha(q)$ and $K_\alpha$ appear in (\ref{8}).

With reference to the conventions (\ref{glnconv}),
we define  the structure constants $c_{\alpha,\beta}^{\alpha+\beta}$
by writing 
$[E_\alpha, E_\beta]=c_{\alpha,\beta}^{\alpha+\beta} E_{\alpha+\beta}$ 
if $\alpha,\beta, (\alpha+\beta)$ all belong to $\Phi$,
and $c_{\alpha,\beta}^{\alpha+\beta}:=0$ otherwise.
Then (\ref{puregauge}) yields 
\bea
&&{\partial}_l A^i_k - {\partial}_k A^i_l = 
\sum_{\alpha\in \Phi} \alpha_i A_l^\alpha A_k^{-\alpha},
\quad
\forall i,k,l , 
\label{Hi}\\
&&
{\partial}_l A_k^\alpha - {\partial}_k A_l^\alpha=
\sum_{i=1}^n  \alpha_i (A^i_l A_k^\alpha - A_k^i A_l^\alpha)
+\sum_{\gamma\in \Phi} c^\alpha_{\gamma, \alpha-\gamma} 
A_l^\gamma A_k^{\alpha-\gamma},
\quad
\forall \alpha,\, \forall k,l.
\label{Ea}\eea
The $H_i\otimes H_j$ and $H_i\otimes E_\alpha$ 
components of (\ref{condonA}) require that  
\bea
&&
{\partial}_k Q^j +{\partial}_j A_k^i +
\sum_{\alpha\in \Phi} \alpha_j r_i^{\alpha} A_k^{-\alpha}=0,
\qquad\forall i,j,k,
\label{HiHj}\\
&& 
{\partial}_k r_i^{\alpha} -\alpha_i F_\alpha A_k^\alpha 
+ \sum_{j=1}^n \alpha_j Q^j A_k^\alpha 
 -\sum_{j=1}^n \alpha_j A_k^j r_i^{\alpha}
+\sum_{\gamma\in \Phi} c_{\gamma, \alpha-\gamma} ^\alpha r_i^{\gamma}
A_k^{\alpha-\gamma} + \sum_{j=1}^n \alpha_j A_j^i A_k^\alpha=0
\qquad{} 
\label{HiEa}\eea
$\forall i,k, \alpha$.
{}From the $E_\alpha\otimes H_i$ and $E_\alpha\otimes E_\beta$ components 
of (\ref{condonA}) we find that 
\bea
&&
{\partial}_i A_k^\alpha + \alpha_i F_\alpha A_k^\alpha=0,
\quad
\forall i,k,\alpha,
\label{EaHi}\\
&&
\delta_{\beta,-\alpha}\alpha_{k} w^{2}_{\alpha} -
c^{\alpha +\beta}_{\alpha ,\beta}
\frac{ w_{\alpha}w_{\beta} }{ w_{\alpha +\beta} } A^{\alpha +\beta}_{k}
+\sum_{j=1}^n\alpha_{j}r_j^{\beta}A^{\alpha}_{k} +
\sum_{j=1}^n \beta_{j}A^{\alpha}_{j} A^{\beta}_{k}=0
\label{EaEb}
\eea
$\forall k, \alpha,\beta$.
Note that to derive (\ref{EaEb}) we have used the identities 
(\ref{derprop}), (\ref{lemma2})
and the symmetry properties of the structure constants.

It is convenient to focus first on the last two
equations, 
since they do not contain the Cartan components of $A_k$.
Eq.~(\ref{EaHi}) obviously implies that
\be
A_k^\alpha(q) = w_\alpha(q) b_k^\alpha, 
\qquad
b_k^\alpha: \hbox{some constants}.
\label{formB}\ee
The constants are then determined as follows.

\smallskip
\noindent
{\bf Proposition 2.}
{\em Eq.~(\ref{EaEb}) admits
solution for the constants $b_k^\alpha$ only for those two families 
of $r(q)$ in (\ref{8}) for which the $C_\alpha$ are chosen according to
\be
\hbox{case I}:\quad C_\alpha = - H_\alpha
\quad \forall \alpha\in \Phi, 
\qquad
\hbox{or}\qquad 
\hbox{case II:}\quad 
C_\alpha= H_\alpha 
\quad \forall \alpha\in \Phi.
\label{Ccases}\ee
For $\alpha=\lambda_m-\lambda_l$,  the $b_k^\alpha$ are 
respectively given by 
\be
 b_k^{(\lambda_m- \lambda_l)}=\delta_{km} + \Omega
\quad 
\hbox{in case I,}\quad\hbox{and}\quad
b_k^{(\lambda_m- \lambda_l)}=\delta_{kl} + \Omega
\quad \hbox{in case II,}
\label{bk}
\label{bcases}\ee
where $\Omega$ is an arbitrary constant.}

\smallskip
\noindent{\bf Proof.} 
The statement is obtained  by an elementary, but rather 
lengthy inspection of eq.~(\ref{EaEb}).
This is contained in appendix B.
{\em Q.E.D.} 

It is easy to explain why we got {\em two} series of solutions 
in the above.
Namely, they arise due to the fact that 
$L$ in (\ref{6}) is a self-adjoint matrix.
Indeed, $L^\dagger =L$ implies that if $r(q)$ solves 
(\ref{2}) then 
$r^\dagger(q)$ also solves it, where 
$(u_1\otimes u_2)^\dagger= u_1^\dagger \otimes u_2^\dagger$.
Furthermore, if $r(q)$ is gauge transformed to a constant
$r'$ 
by $g(q)$, then $r^\dagger(q)$ is transformed 
to  $(r')^\dagger$  by $(g^\dagger)^{-1}$.
The two series of solutions described in Proposition 2 
are exchanged by this symmetry.
It is thus enough to consider only one of these series,
and from now on we concentrate on case I.

As the main result of this section, we now give the 
most general `gauge potential' $A_k$ and $r(q)$ for which $r'$ (\ref{Rprime}) 
will be constant. 

\smallskip
\noindent
{\bf Theorem 3.}
{\em
The most general solution of  
eqs.~(\ref{puregauge}), (\ref{condonA}) for $A_k$ and $Q$ 
in case I of Proposition 2 can be described as follows.
The root part of $A_k$ is determined by Proposition 2, 
while its  Cartan part has the form\footnote{Note that 
$F_{\lambda_l -\lambda_l}=0$ by the 
definition of $F_{\lambda_l-\lambda_k}$ 
in (\ref{f-sub}).}
\be
A_k^l=F_{\lambda_l - \lambda_k} + \Omega \sum_{m\, (m\neq l)} 
F_{\lambda_l - \lambda_m} +{\partial}_k \theta  
\qquad
(\forall k,l=1,\ldots, n),
\label{solPsi}\ee
where $\theta(q)$ is arbitrary smooth 
function.
The function  $Q(q)\in gl_n$ is given by  
\be
Q= -\sum_{k=1}^n A_k^k H_k - 
\Omega \sum_{\alpha\in \Phi}  w_\alpha E_\alpha + g^{-1} Q' g,
\label{solQ}\ee
where  $g(q)\in GL_n$ denotes a solution of ${\partial}_k g= - g A_k$ and 
$Q'\in gl_n$ is an arbitrary constant.
}

\smallskip
\noindent{\bf Proof.} 
The main steps of the proof can be outlined as follows.
After choosing case I of Proposition 2, the 
right hand side of (\ref{Hi}) can be calculated.
The general solution of (\ref{Hi}) for the unknowns 
$A_k^l$ is then found to be
\be
A_k^l=F_{\lambda_l - \lambda_k} + \Omega \sum_{m\, (m\neq l)} 
F_{\lambda_l - \lambda_m} +{\partial}_k \theta^l  
\qquad
(\forall k,l=1,\ldots, n),
\label{solHi}\ee
where the $\theta^l$ are arbitrary smooth functions of $q$.
Next, it is verified that (\ref{solHi})  solves (\ref{Ea})
if and only if
\be
\theta^1=\theta^2=\cdots = \theta^n:=\theta.
\label{solEa}\ee
At this point we have the general solution for $A_k$ and
remaining task is to solve (\ref{HiHj}), (\ref{HiEa})
for $Q$. By using also (\ref{ria}) with $C_\alpha = -H_\alpha$,
these are inhomogeneous linear differential equations for $Q$. 
It is an easy matter to check    
that (\ref{solQ}) with $Q'=0$
gives a particular solution, and that the difference $\delta Q$ 
of two solutions must satisfy the equations
\be
{\partial}_k (\delta Q) + [\delta Q, A_k]=0
\qquad
(\forall k=1,\ldots, n).
\ee
The proof is completed by remarking that the last equation 
is equivalent to  ${\partial}_k (g (\delta Q)  g^{-1}) =0$ 
with ${\partial}_k g = -g A_k $.
{\em Q.E.D.}

We wish to make some observations  on the above result.
Firstly,  note that if $r'$ is the constant r-matrix 
obtained from (\ref{Rprime}) in the case  
\be
\theta = 0,
\qquad
Q'=0,
\label{setto0}\ee
then in the general case of Theorem 3 the same formula  yields
\be
r' + {\bf 1}_n \otimes Q'.
\ee
This means that the free parameters $\theta$ and $Q'$ 
in (\ref{solPsi}), (\ref{solQ}) are irrelevant. 
Henceforth they will be set to zero.
An additional convenience  of this choice 
is that it guarantees the antisymmetry 
of $r'$ (\ref{Rprime}).
In fact, one can compute the symmetric part of
$(r + \sum_k A_k \otimes H_k)$ and finds it to be zero if $Q'=0$.
Secondly,  it is worth pointing out that 
\be
r'\in sl_n \wedge sl_n
\qquad
\Leftrightarrow \qquad 
\Omega = -{1\over n}.
\label{slncond}\ee
Indeed, the condition $r'\in sl_n\wedge sl_n$ is clearly 
equivalent to 
$(r + \sum_k A_k \otimes H_k)\in sl_n \wedge sl_n$,
and this is easily calculated to hold if and only if
$Q'=0$ and $\Omega=-{1\over n}$.
Since for a given $A_k$ 
the solution of (\ref{defA}) for $g(q)\in GL_n$ is 
unique up to a constant, 
\be
g(q) \rightarrow g_0 g(q),
\qquad
\forall
g_0 \in GL_n,
\ee
we can also conclude that 
if the condition $r'\in sl_n \otimes sl_n$ is imposed,
then $r'$ is necessarily  
antisymmetric and is uniquely determined up to an automorphism of $sl_n$.

Finally, let us observe that 
our $r(q)$ and  $A_k(q)$ 
for which $r'$ will be a constant 
admit the interesting decompositions 
\be
r= \tilde r - \Omega  {\bf 1}_n \otimes {\cal A} ,
\qquad 
A_k = \tilde A_k   + \Omega {\cal A},
\label{A-decomp}\ee
where 
\be
{\cal A}= \sum_{l,m\, (l\neq m)} \left( F_{\lambda_l - \lambda_m} H_l
+ w_{\lambda_l -\lambda_m} E_{\lambda_l - \lambda_m}\right). 
\label{calA}\ee
Here $r(q)$, $A_k$ are given  by Theorem  3 together with 
(\ref{setto0}).
In the rest of the paper we shall determine the corresponding 
constant r-matrices from (\ref{Rprime}).
It will be convenient to consider first  
the $\Omega=0$ special case,
for which $r$, $A_k$, $r'$ become $\tilde r$,  $\tilde A_k$, $\tilde r'$, 
respectively. 

\section{Constant r-matrices  from gauge transformation}

If $A_k$ is given so that (\ref{puregauge}) holds 
then the gauge transformation $g(q)$ can be determined from 
the differential equation in (\ref{defA}). 
By taking $A_k$ and $r(q)$ from Theorem 3 with (\ref{setto0}), 
this $g$ will transform the 
dynamical r-matrix $r(q)$ into an antisymmetric constant (\ref{Rprime}). 
Here we shall determine $g(q)$ and $r'$ explicitly.
For an antisymmetric constant $r'$ 
the (modified) classical Yang-Baxter equation
is a sufficient condition for 
the Jacobi identity 
$\{\{ L'_1, L'_2\}, L'_3\} + {\mathrm{cycl.}}=0$,
which will be seen to hold for 
the r-matrices found below.

\subsection{The case of $\Omega=0$}

Now we calculate the gauge transformation and 
the  resulting constant r-matrix 
in the special case of Theorem 3 for which $\Omega=0$
and (\ref{setto0}) hold.
In agreement with (\ref{A-decomp}), the various quantities 
will carry a tilde in this case.
We shall use the notation
\be
I^n_k:= \{ 1,\ldots,n\} \setminus \{ k\},
\qquad 
\forall k=1,\ldots,n,
\label{I-set}\ee
and write the elements of $gl_n$  as matrices.
Then  $\tilde r(q)$ and $\tilde A_k(q)$ take the following form: 
\be
\tilde r=-\sum_{1\leq k\neq l\leq n}\left(F_{kl}
e_{kl}\otimes e_{lk}
+ w_{kl} e_{kk}\otimes e_{kl}\right),
\quad
\tilde A_k = \sum_{l\in I^n_k}\left( w_{kl} e_{kl} +
F_{lk} e_{ll}\right).    
\label{RA}\ee

Let us start by defining the matrix function $\varphi$ of $q$ as follows:
$\varphi_{nk}:=1$ for any $k=1,\ldots, n$ and 
\be
\varphi_{jk}:=
\sum_{
\begin{array}{c} 
P\subset I^n_k \\ \vert P\vert=n-j
\end{array}}
\left( \prod_{l\in P} F_l \right) 
\qquad\qquad
\forall k, \quad 
1\leq j \leq n-1,
\label{varphi}\ee
where $\vert P \vert$ denotes the number of the elements of $P$.
Moreover, let $\chi$ be the $n\times n$ matrix 
function of $q$ given by
\be
\chi_{jk} = \delta_{jk} \prod_{l\in I_k^n} {1\over  w_l}.
\label{chi}\ee
These formulae yield invertible matrices on 
the admissible domain of $q$, where $v(q)$ is finite.
This is obvious for the diagonal matrix $\chi$.
By using the identity
\be
\sum_{l=1}^n (-F_i)^{l-1} \varphi_{lj}= \prod_{\tau\in I_j^n} (F_\tau-F_i),
\ee 
we can also find the inverse of $\varphi$ explicitly
\be
\left(\varphi^{-1}\right)_{jk}=  
(-F_j)^{k-1} \prod_{l\in I_j^n} {1\over (F_l - F_j)}.
\ee  

\smallskip
\noindent
{\bf Proposition 4.}
{\em A gauge transformation $\tilde g(q)\in GL_n$ that satisfies 
\be
{\partial}_k \tilde g(q)=- \tilde g(q)\tilde A_k(q)
\label{tilgdiff}\ee
with $\tilde A_k$ in (\ref{RA}) is given  by 
$\tilde g(q)=\varphi(q) \chi(q) $, where $\varphi$ and $\chi$ are defined 
by (\ref{varphi}) and (\ref{chi}).}

\smallskip
\noindent{\bf Proof.} 
The componentwise form of (\ref{tilgdiff})
with $\tilde A_k$ in (\ref{RA}) reads
\be
{\partial}_k \tilde g_{ik}=0, \qquad \forall i,k\in \{ 1,\ldots, n\},
\label{*1}\ee
\be
{\partial}_k \tilde g_{ij}= 
- \tilde g_{ij} F_{jk} -\tilde  g_{ik} w_{kj}, \qquad
\forall i,j,k\in \{ 1,\ldots, n\},\quad j\neq k.
\label{*2}\ee
We  notice that the matrix 
\be
\tilde g_{ij}(q)= \prod_{l\in I_j^n} {1\over w(q_l + c_i)},
\qquad 
i,j\in \{1,\ldots, n\},
\label{*g}\ee
where the $\{ c_i\}_{i=1}^n$ are pairwise distinct constants,
yields a solution of these equations.
Indeed, (\ref{*1}) holds obviously, 
while (\ref{*2}) is  checked with the aid of the identity (\ref{lemma2}).
Using (\ref{lemma2}) again, we can rewrite the matrix $\tilde g(q)$ 
defined by (\ref{*g}) 
in the product form    
\be
\tilde g(q) ={\mathbf C} \varphi(q) \chi(q),
\ee
where ${\mathbf C}$ is the invertible constant matrix given by 
\be
{\mathbf C}_{ij} = {1\over w(c_i)^{n-1}} \left( F(c_i)\right)^{j-1}.
\ee
Since equation 
(\ref{tilgdiff}) determines $\tilde g$ up to multiplication by
a constant matrix form the left, the required statement follows. {\em Q.E.D.} 

We can now calculate the gauge transformed r-matrix from (\ref{Rprime}).
The result turns 
out to be an antisymmetric, constant 
solution of the (modified) classical Yang-Baxter equation,
\be
[r'_{12}, r'_{13}] + [r'_{12},  r'_{23}] + 
[r'_{13}, r'_{23}]=- {\cal B} \hat {\cal F},
\label{CYB}\ee
where ${\cal B}$ appears in (\ref{C}) and 
$\hat {\cal F}\in (gl_n)^{3\wedge}$ is given by
\be
\hat {\cal F}:= \sum_{i,j,k,l,r,s=1}^n {\cal F}_{ij,kl}^{rs} 
e_{ji}\otimes e_{lk} \otimes e_{rs}
\quad\hbox{with}\quad
[e_{ij}, e_{kl}]=\sum_{r,s=1}^n {\cal F}_{ij,kl}^{rs} e_{rs}.
\label{hatF}\ee

\smallskip
\noindent
{\bf Proposition  5.}
{\em The gauge transform of $\tilde r(q)$ in (\ref{RA}) by $\tilde g(q)$ 
in Proposition 4 
is given by  
\be
\tilde r'= \sum_{(a,b,c,d)\in S}\,
 \left( {\cal B} e_{ab}\wedge e_{cd} - 
e_{a+1,b}\wedge e_{c+1,d}\right),
\label{constR}\ee
$$
S=\{\, (a,b,c,d)\in {\mathbf N}^4 \,\vert \,
a+c+1=b+d,
\quad 
1\leq b\leq a <n,
\quad
b\leq c <n,
\quad 
1\leq d\leq n \,\}.
$$
This formula defines an antisymmetric solution of (\ref{CYB}). 
}
\smallskip

\smallskip
\noindent
{\bf Proof.}
The first statement is verified by a direct calculation, 
which is described in appendix C.
The fact that $\tilde r'$ solves (\ref{CYB}) can also be checked directly.
Alternatively, it follows from the identification of $\tilde r'$ 
in terms of certain well-known solutions of (\ref{CYB}),
which is presented in section 5. {\em Q.E.D.}

It is clear from (\ref{CYB}) that the two terms in (\ref{constR}) must 
separately satisfy the classical Yang-Baxter equation,
\be
[b_{12}, b_{13}] + [b_{12}, b_{23}] + [b_{13}, b_{23}]=0.
\label{YB0}\ee
In fact, this holds since the first term 
\be
b_{gl_n}:=\sum_{(a,b,c,d)\in S} e_{ab}\wedge e_{cd}
\label{firstb}\ee
is nothing but the classical r-matrix associated with the Frobenius subalgebra 
of $gl_n$ spanned by the matrices with vanishing last row, 
which is described as an example in \cite{BD}.
More explicitly, it reads as   
\be
b_{gl_n}=\sum_{k=1}^{n-1} \sum_{j=1}^{n-k} e_{jj}\wedge e_{n-k, n+1-k}
+\sum_{1\leq i<j\leq n} \sum_{m=1}^{j-i-1} 
e_{n+1-i-m,n+1-j} \wedge e_{n+m-j,n+1-i}.
\label{bgln}\ee
The second term is a transform of the first one according to 
\be
\sum_{(a,b,c,d)\in S} e_{a+1,b}\wedge e_{c+1,d}= 
-(\sigma \otimes \sigma)b_{gl_n},
\ee
where  
$\sigma: gl_n \rightarrow gl_n$ is the inner automorphism 
\be
\sigma: e_{ij} \mapsto e_{n+1-i, n+1-j}.
\label{sigma}\ee
Finally, we note for later purpose that
\be
\tilde r' = {\cal B} b_{gl_n} + (\sigma\otimes \sigma) b_{gl_n}=
 \tilde r'_{sl_n} + X \wedge {\bf 1}_n,
\label{rX}\ee
where $\tilde r'_{sl_n}\in sl_n \wedge sl_n$ and
\be
X=-{1\over n} 
\sum_{k=1}^{n-1} (n-k) e_{k+1, k}
-{{\cal B}\over n}  \sum_{k=1}^{n-1} k e_{k, k+1}.
\label{X}\ee
Of course, $\tilde r'_{sl_n}$ satisfies the same equation
(\ref{CYB}) as $\tilde r'$. 

\subsection{The case of an arbitrary $\Omega$}

Now we tackle the general case by 
making use of the decompositions of  $r(q)$ and $A_k$ in (\ref{A-decomp}).

It is natural to look for $g(q)$ as a product
\be
g(q)= h(q) \tilde g(q),
\ee
where $\tilde g(q)$ is given in Proposition 4.
Then the equation ${\partial}_k g = - (\tilde A_k + \Omega {\cal A})g$
reduces to  
\be
{\partial}_k h= - h {\tilde {\cal A}}\Omega
\quad\mbox{with}\quad
\tilde {\cal A}:= \tilde g {\cal  A} {\tilde g}^{-1},
\label{hdiff}\ee
where ${\cal A}$ is given in (\ref{calA}).
By using also the decomposition of $r(q)$ in (\ref{A-decomp})
we obtain from (\ref{Rprime}) that
\be
r' = (h(q)\otimes h(q)) \left(\tilde r' + 
\Omega \tilde {\cal A}(q) \wedge {\bf 1}_n \right) 
(h(q) \otimes h(q))^{-1},
\label{hr}\ee
where $\tilde r'$ is given by (\ref{constR}).
The fact that $r'$ and $\tilde r'$ are  both constant 
permits  us to prove the following result without further explicit
calculation.

\smallskip
\noindent
{\bf Theorem 6.}
{\em With the above notations and $\tilde r'$, $X$ defined in 
(\ref{constR}), (\ref{rX}), 
we have
\be
h(q)= g_0\exp\left(- X n\Omega  \sum_{i=1}^n q_i \right), 
\label{hq}\ee
where $g_0\in GL_n$ is an arbitrary constant,
and 
\be
r'= (g_0 \otimes g_0)
\left( \tilde r'_{sl_n} + (n\Omega  +1) X \wedge {\bf 1}_n \right)
(g_0 \otimes g_0)^{-1}
\label{r'gen}\ee
is the most general constant r-matrix resulting from gauge transformation.}

\smallskip
\noindent
{\bf Proof.}
By substituting (\ref{rX}), we can rewrite (\ref{hr}) 
as the sum
$r' = r'_{sl_n} + r'_{\mathrm rest}$
with
\be
r'_{sl_n} = 
(h(q)\otimes h(q)) \tilde r'_{sl_n}(h(q) \otimes h(q))^{-1}
\label{r'sln}\ee
and 
\be
r'_{\mathrm rest}=
\left(h(q)( \Omega \tilde {\cal A}(q) + X ) h^{-1}(q)\right) \wedge {\bf 1}_n.
\label{r'rest}\ee
Since $r'$ is constant, these two terms must be constant separately.
Recall now that $\tilde {\cal A}(q)$ is independent of $\Omega$
by its definition (\ref{hdiff}) and that for $\Omega=-{1\over n}$ we must
have $r'\in sl_n \wedge sl_n$ (\ref{slncond}).
This implies that $(X-{1\over n}\tilde {\cal A}(q))$ must vanish, whereby 
\be
\tilde  {\cal A} = n X.
\label{calA=}\ee
Hence we obtain (\ref{hq}) from the differential equation in (\ref{hdiff}).
But then the fact that $r'_{sl_n}$ is constant shows that
the relation
\be 
[X \otimes {\bf 1}_n + {\bf 1}_n \otimes X, \tilde r'_{sl_n}]=0,
\label{Xr}\ee
which is equivalent to 
\be
r'_{sl_n}= (g_0 \otimes g_0) {\tilde r}'_{sl_n} 
(g_0 \otimes g_0)^{-1},
\ee
must be valid.
By substituting these results back into (\ref{hr}) we arrive at (\ref{r'gen}). 
{\em Q.E.D.}

Incidentally, we have also verified by explicit calculation
that (\ref{calA=}) and (\ref{Xr}) are indeed
satisfied, which represents a reassuring check on the foregoing
considerations in the paper.

\section{Identification of the constant r-matrices}

The constant r-matrix (\ref{r'gen}) is a solution
of (\ref{CYB}).
For the rational Calogero-Moser model, ${\cal B}=0$, this is 
the classical Yang-Baxter equation.
In this case the identification 
of the r-matrix in terms of a Frobenius subalgebra of $gl_n$ 
has already been mentioned (\ref{rX}). 
In the hyperbolic/trigonometric cases (\ref{CYB}) 
is  the modified classical Yang-Baxter equation, whose
antisymmetric solutions have been classified by 
Belavin and Drinfeld \cite{BD} for the complex
simple Lie algebras. 
A well-known solution for the Lie algebra $sl_n$,
with the normalization  
\be
[\rho_{12}, \rho_{13}] + [\rho_{12}, \rho_{23}] + 
[\rho_{13}, \rho_{23}]= - \hat {\cal F}, 
\label{CYB-1}\ee
is the so-called Cremmer-Gervais classical r-matrix, 
which we quote from \cite{GG} as 
\be
r_{CG}= \sum_{1\leq i<j\leq n} e_{ij}\wedge e_{ji} 
+2 \sum_{1\leq i<j\leq n} \sum_{m=1}^{j-i-1} e_{i,j-m} \wedge e_{j,i+m}
+{1\over n} \sum_{1\leq i<j\leq n} (n+2(i-j)) e_{ii}\wedge e_{jj}.
\label{rCG}\ee
Note that $r_{CG}\in sl_n \wedge sl_n$ 
and $\hat {\cal F}$ (\ref{hatF}) belongs to $(sl_n)^{3 \wedge}$. 
Below we  show that for ${\cal B}\neq 0$ the $sl_n$-part of the 
constant Calogero-Moser r-matrix (\ref{r'gen}) is equivalent to $r_{CG}$.

We shall need the following  
properties of $r_{CG}$.
As in \cite{GG}, first introduce $J_0, J_\pm \in sl_n$ by 
\be
J_0= {1\over 2} \sum_{k=1}^n  (n+1 - 2k) e_{kk},
\qquad
J_+ = \sum_{k=1}^{n-1} (n-k) e_{k, k+1},
\qquad
J_- = \sigma(J_+)= \sum_{k=1}^{n-1} k e_{k+1, k}.
\label{Js}\ee 
They generate the principal $sl_2$ subalgebra of $sl_n$,
\be
[J_0, J_\pm] = \pm J_\pm,
\qquad
[J_+, J_-]=2J_0.
\label{sl2}\ee
Then define the elements 
$b_{CG}^\pm := \mp {1\over 2}
[ J_\pm \otimes {\bf 1}_n + {\bf 1}_n \otimes J_\pm, 
r_{CG}]\in sl_n\wedge sl_n$.
Explicitly, 
\be
b_{CG}^+= \sum_{k=1}^{n-1} d_k \wedge e_{k,k+1} + 
\sum_{1\leq i<j\leq n} \sum_{m=1}^{j-i-1}
e_{i,j-m+1} \wedge e_{j,i+m},
\qquad\quad
d_k:=  \sum_{j=1}^k e_{jj} -{k\over n} {\bf 1}_n.
\label{bcg+}\ee
On account of  $(\sigma \otimes \sigma) r_{CG}=-r_{CG}$,
with $\sigma$ defined in (\ref{sigma}),
$b_{CG}^-= (\sigma\otimes \sigma) b_{CG}^+$.
It has been pointed out in \cite{GG} that the subspace of $sl_n \wedge sl_n$ 
spanned by $r_{CG}$ and $b^\pm_{CG}$ is an irreducible 
representation of the principal $sl_2$ subalgebra. 
In fact, for the operators
\be
{\cal J}_{0,\pm} (Y):= [ J_{0,\pm} \otimes {\bf 1}_n + 
{\bf 1}_n\otimes J_{0,\pm}, Y]
\qquad
\forall Y\in gl_n\otimes gl_n,
\ee
one has the relations:
\be
{\cal J}_0 \left( \begin{array}{c} b_{CG}^+ \\ r_{CG} \\ 
b_{CG}^- \end{array} \right)
=\left( \begin{array}{c} b_{CG}^+ \\ 0 \\ -b_{CG}^- \end{array} \right),
\quad
{\cal J}_+ \left( \begin{array}{c} b_{CG}^+ \\ r_{CG} \\ b_{CG}^- 
\end{array} \right)
=\left( \begin{array}{c} 0 \\ -2 b_{CG}^+ \\  r_{CG} \end{array} \right),
\quad
{\cal J}_- \left( \begin{array}{c} b_{CG}^+ \\ r_{CG} \\ b_{CG}^- 
\end{array} \right)
=\left( \begin{array}{c} -r_{CG} \\ 2 b_{CG}^- \\ 0 \end{array} \right).
\label{sl2rep}\ee
It follows from these relations  that $b_{CG}^\pm$ satisfy 
the classical Yang-Baxter equation \cite{GG}, and 
the identification of $b_{CG}^\pm$ in terms of 
Frobenius subalgebras of $sl_n$ is also described in this reference.

Now we are prepared to establish the connection 
between $r_{CG}$ and the r-matrix $r'$ (\ref{r'gen}). 
The key observation is the following identity:
\be
- (T\otimes T)  \tilde r'_{sl_n} = b_{CG}^+ + {\cal B} b_{CG}^-,
\label{keyrel}\ee
where $T: gl_n \rightarrow gl_n$ denotes matrix transposition.
This can be checked directly from the formulas (\ref{rX}), 
(\ref{bgln}) (\ref{bcg+}).
It permits us to transform $\tilde r'_{sl_n}$ into a multiple of $r_{CG}$ 
in a simple manner.
To treat the hyperbolic/trigonometric cases together,
we introduce  the parameter
\be
a'=\left\{
\begin{array}{cc} 
a,&\mbox{hyperbolic case,}\\
\sqrt{-1}a, &\mbox{trigonometric case,}\end{array}
\right.
\label{mu}\ee
whose square ${\cal B}=(a')^2$ appears in (\ref{CYB}).
By using (\ref{sl2rep}) it is not difficult to verify that 
\be
(u_- u_+ \otimes u_- u_+) \left(T\otimes T \tilde r'_{sl_n} \right)  
( u_- u_+ \otimes u_- u_+)^{-1} = a' r_{CG}
\label{hyptraf}\ee
with 
\be
u_-:= \exp\left({a'\over 2} J_-\right),
\qquad 
u_+:= \exp\left(-{1\over a'}  J_+\right).
\ee
According to (\ref{hyptraf}) the $sl_n$-part of $r'$
is equivalent to $a'r_{CG}$ under a Lie algebra automorphism. 

In the end, 
notice  from (\ref{X}) and (\ref{Js})  that 
\be
X=- {1\over n} ( J_+^T + {\cal B} J_-^T).
\label{Xsl2}\ee 
This allows us to present the r-matrix associated with
\be
L'(q,p) = g_0 h(q) \tilde g(q) L(q,p) (g_0 h(q) \tilde g(q))^{-1} 
\ee
in a `standard form'.
Here $h(q)$ and $\tilde g(q)$ are the same as in Theorem 6,
and our final result is formulated as follows.

\smallskip
\noindent
{\bf Proposition  7.}
{\em Consider the hyperbolic/trigonometric Calogero-Moser models.
If in Theorem 6  the constant $g_0$ is chosen to be 
\be
g_0= \exp\left(-{a'\over 2} J_-^T\right)
 \exp\left({1\over a'}  J_+^T\right),
\ee
then the r-matrix (\ref{r'gen}) becomes
\be
r'= a' (T\otimes T) (r_{CG} + 2(\Omega + {1\over n}) J_0 \wedge {\bf 1}_n ).
\ee
}
\smallskip
\noindent
{\bf Proof.}
By means of the 
$sl_2$ algebra (\ref{sl2})  and (\ref{Xsl2})
it is easy to check that $g_0 X g_0^{-1} = {2a'\over n} J_0^T$.
The statement is obtained by
combining this with (\ref{hyptraf}).
{\em Q.E.D.}

This proposition describes the precise relationship 
between the most general constant r-matrices of the hyperbolic/trigonometric 
Calogero-Moser models and the standard Cremmer-Gervais 
classical r-matrices.  

\section{Conclusion}

In this paper we have determined the most general constant
r-matrices that may be obtained by coordinate dependent gauge transformations 
of the standard Lax representation (\ref{6}) of the degenerate Calogero-Moser
models associated with $gl_n$.
Up to automorhisms of $gl_n$ 
(i.e. up to conjugation by constants $g_0\in GL_n$ and transpose)
and addition 
of an irrelevant term ${\bf 1}_n \otimes Q'$ with any constant $Q'\in gl_n$,
the most general such r-matrix turned out to have the form 
\be
r'= \sum_{(a,b,c,d)\in S}\,
 \left( {\cal B} e_{ab}\wedge e_{cd} - 
e_{a+1,b}\wedge e_{c+1,d}\right) 
+n\Omega X
 \wedge {\bf 1}_n,
\label{+constR}\ee
where 
$$
X=-{1\over n}\sum_{k=1}^{n-1} (n-k) e_{k+1, k}
-{{\cal B}\over n} \sum_{k=1}^{n-1} k e_{k, k+1},
$$
${\cal B}$ is given according to (\ref{C}) 
in correspondence with the rational, hyperbolic and trigonometric 
potential functions (\ref{4}), 
$\Omega$ is an arbitrary constant scalar, and   
$$
S=\{\, (a,b,c,d)\in {\mathbf N}^4 \,\vert \,
a+c+1=b+d,
\quad 
1\leq b\leq a <n,
\quad
b\leq c <n,
\quad 
1\leq d\leq n \,\}.
$$ 
We have seen that $r'$ solves the classical (modified) Yang-Baxter  equation
(\ref{CYB}), and have identified it 
in terms of well-known solutions of this equation.
In particular, we have shown that in the hyperbolic and trigonometric
cases the above $r'$ with $\Omega=-{1\over n}$ is equivalent to a multiple of the
Cremmer-Gervais classical r-matrix under an automorphism of $gl_n$.
We obtained 
these results by an explicit determination of the  
gauge transformations $g(q)\in GL_n$ for which the Poisson brackets
of  
$L'(q,p)= g(q) L(q,p) g^{-1}(q)$, where $L$ is the standard Lax matrix 
(\ref{6}), can be written in the form (\ref{2})
with a constant r-matrix.  
The gauge transformation $g(q)$ for which the Poisson brackets 
of $L'$ are encoded by $r'$ in (\ref{+constR}) was found as the product 
\be
g(q)=\exp\left( -X n\Omega \sum_{i=1}^n q_i \right)\varphi(q)\chi(q), 
\ee
where the matrices $\varphi(q)$ and $\chi(q)$ are 
defined by (\ref{varphi}) and (\ref{chi}), 
with the notations  
fixed by equations (\ref{7}), (\ref{F}), (\ref{f-sub}) in section 2.   

The outcome of our direct analysis of the degenerate 
Calogero-Moser models is consistent 
with the results obtained in \cite{AHZ,H,Hou} by different means.
We hope to present a more detailed comparison with the
elliptic case as well as an analogous study for other Lie algebras
elsewhere.

\bigskip
\bigskip
\noindent
{\small 
{\bf  Acknowledgments.}
We wish to thank J. Balog and the referees for useful comments on 
the manuscript. This work has been supported in part by the 
Hungarian Ministry  of Education under FKFP 0596/1999 and by the 
National Science Fund (OTKA) under T025120, M028418. }
\bigskip

\newpage
\renewcommand{\theequation}{\arabic{section}.\arabic{equation}}
\renewcommand{\thesection}{\Alph{section}}
\setcounter{section}{0} 

\section{Proof of Theorem 1}
\setcounter{equation}{0}
\renewcommand{\theequation}{A.\arabic{equation}}

The proof given below relies on
the general analysis of the momentum independent Calogero-Moser 
r-matrices presented by Braden and Suzuki in \cite{BS}.
We first specialize the relevant results of \cite{BS} to our case 
and then further elaborate them to obtain the statement of Theorem 1.
   
Consider the Lax matrix in (6) with a function $w$ in (7).
Our task is to find the most general momentum independent r-matrix,
$r(q)$, which satisfies equation (2), i.e.,
\be
\{ L_1, L_2\}(q,p)= [ r_{12}(q), L_1(q,p)] - [r_{21}(q), L_2(q,p)].
\label{2ujra}\ee
Obviously, $r(q)=r_{12}(q)\in gl_n\otimes gl_n$ can be 
expanded in the form
\be
r(q)= \sum_{i,j=1}^n r^{i,j}(q) H_i \otimes H_j 
+ \sum_{\alpha\in \Phi} \sum_{i=1}^n \left( 
r^{i,\alpha}(q) H_i \otimes E_\alpha + 
r^{\alpha,i}(q) E_\alpha \otimes H_i \right)
+ \sum_{\alpha,\beta \in \Phi} 
r^{\alpha,\beta}(q) E_\alpha\otimes E_\beta.
\label{C1}\ee
Since the functions $w$ in (7) are odd 
(and thus $w_{-\alpha}(q)=-w_\alpha(q)$), 
we can use the results of the
third and fourth chapters of [12], where it 
has been shown that under our conditions 
the following equations hold:
\be
r^{\alpha,i}(q)=0\qquad (\forall i\in 
\{1,\ldots,n\},\, \forall \alpha\in \Phi ),
\label{C2}\ee
\be 
r^{\alpha, \beta}(q)= 
\frac{w'_{\alpha}(q)}{w_\alpha(q)} \delta_{\alpha,-\beta}
=- F_\alpha(q) \delta_{\alpha,-\beta}
\qquad (\forall \alpha,\beta \in \Phi).
\label{C4}\ee
Moreover, according to  \cite{BS},  the remaining 
requirements on $r(q)$ reduce to the equations 
\be
\sum_{i=1}^n \alpha_i r^{i,j}(q)=0
\qquad (\forall \alpha\in \Phi,\, \forall j\in \{1,\ldots,n\}),
\label{C5}\ee
and 
\be
\sum_{i=1}^n (\alpha_i r^{i,\beta} w_\alpha - 
\beta_i r^{i,\alpha}  w_\beta) 
=c_{-\alpha, \alpha+\beta}^{\beta} 
(r^{-\alpha,\alpha} w_{\alpha+\beta} +
 r^{-\beta, \beta}
w_{\alpha+\beta} ) \qquad (\forall \alpha,\beta \in \Phi).
\label{C6}\ee
We here use the basis of $gl_n$ introduced in (5),
$\alpha_i:=\alpha(H_i)$, 
the structure constants 
$c_{\alpha,\beta}^{\alpha + \beta}$  satisfy 
$[E_\alpha, E_\beta]=
c_{\alpha,\beta}^{\alpha + \beta} E_{\alpha +\beta}$ if
$\alpha, \beta, (\alpha + \beta)$ all belong to $\Phi$, and
$c_{\alpha,\beta}^{\alpha + \beta}:=0$ otherwise.

Now consider equation (\ref{C5}) for 
$\alpha := (\lambda_k -\lambda_l)\in \Phi$. From this we see that 
 $r^{k,j}(q)- r^{l,j}(q)=0$ $(k\neq l, \forall j)$,
which means that the general solution of (\ref{C5}) is
\be
r^{i,j}(q)= M^j(q)
\qquad
(\forall i,j\in \{ 1,\ldots,n\}),
\label{C7}\ee
where the $M^j$ are arbitrary smooth functions of $q$.
Let us next solve (\ref{C6}) for $r^{i,\alpha}(q)$.
By substituting (\ref{C4}) into (\ref{C6}) and using the identity (10)
and the symmetry properties of the structure constants we obtain
\be
\sum_{i=1}^n \left( \alpha_i \hat r^{i,\beta}(q) -
\beta_i \hat r^{i,\alpha}(q)\right) = c_{\alpha,\beta}^{\alpha + \beta}
\qquad (\forall \alpha,\beta \in \Phi),
\label{C8}\ee
where we define
$\hat r^{i,\gamma}:= \frac{r^{i,\gamma}}{w_\gamma}$ 
for any $\gamma\in \Phi$.
By introducing the notations
\be
\hat r^\alpha_S:= \sum_{i=1}^n 
( \hat r^{i,\alpha}+ \hat r^{i,-\alpha}) H_i,
\qquad
\hat r^\alpha_A:= \sum_{i=1}^n 
( \hat r^{i,\alpha}- \hat r^{i,-\alpha}) H_i,
\ee
we have
\be
\hat r^\alpha:= \sum_{i=1}^n \hat r^{i,\alpha} H_i= 
\frac{1}{2}(\hat r_S^\alpha + \hat r_A^\alpha),
\quad \hat r_S^\alpha(q)= \hat r_S^{-\alpha}(q),\quad
\hat r_A^{-\alpha}(q)= -\hat r_A^{\alpha}(q).
\label{C8.5}\ee
We now consider equation (\ref{C8}) for the pairs of 
roots $(\alpha,\beta)$ and $(\alpha, -\beta)$. 
By adding these two equations we get
\be
\alpha( \hat r_S^\beta(q)) = c_{\alpha,\beta}^{\alpha + \beta} 
+ c_{\alpha, -\beta} ^{\alpha - \beta} 
\qquad (\forall \alpha, \beta \in \Phi).
\label{C9}\ee
It follows from the definition of $K_\alpha$ in (5) that
$\alpha(K_\beta)=-(c_{\alpha,\beta}^{\alpha + \beta} 
+ c_{\alpha, -\beta} ^{\alpha - \beta})$ 
for any $\alpha, \beta \in \Phi$.
Therefore the general solution of (\ref{C9}) is given by
\be
\hat r_S^\alpha (q)= - K_\alpha + \tau^\alpha_S(q) {\mathbf 1}_n 
\qquad (\forall \alpha \in \Phi),
\label{C11}\ee
where $\tau^\alpha_S(q)= \tau^{-\alpha}_S(q)$ are 
arbitrary smooth functions.
On the other hand,
by substituting (\ref{C9}) and  
the decomposition in (\ref{C8.5}) into (\ref{C8})  we 
obtain the relation
\be
\alpha( \hat r_A^\beta(q)) = \beta( \hat r_A^\alpha(q))
\qquad (\forall \alpha,\beta \in \Phi).
\label{C10}\ee
Obviously, there exists the decomposition
\be
\hat r^\alpha_A(q) = C_\alpha(q) + \tau_A^\alpha(q) {\mathbf 1}_n,
\label{C12}\ee
where $C_\alpha(q)\in {\cal H}_n\subset sl_n$ and $\tau_A^\alpha(q)$ 
are smooth functions.
The antisymmetry of $\hat r_A^\alpha(q)$ in $\alpha$ and
(\ref{C10}) can be rewritten as 
\be
C_{-\alpha}(q)=-C_\alpha(q),
\qquad
\alpha (C_\beta(q)) = \beta (C_\alpha(q)),
\qquad
\tau_A^{-\alpha}(q)= -\tau_A^\alpha(q).
\label{C13}\ee

By the above, we have parametrized the most general $r(q)$ in terms 
of the functions $M^j$, $\tau^\alpha_A$, $\tau_S^\alpha$ and $C_\alpha$.
If we now introduce the notation
\be
Q(q):= \sum_{i=1}^n M^i(q) H_i + 
\frac{1}{2} \sum_{\alpha\in \Phi} 
\left( \tau^\alpha_S(q) + 
\tau_A^\alpha(q)\right) w_\alpha(q) E_\alpha,
\label{C14}
\ee
then $r(q)$ in (\ref{C1}) takes precisely 
the form stated by Theorem 1, which completes the proof.

\newpage
\section{Proof of Proposition 2}
\setcounter{equation}{0}
\renewcommand{\theequation}{B.\arabic{equation}}

In this appendix we prove Proposition 2  by analyzing equation 
(\ref{EaEb}),
\be
\alpha_{k} w^{2}_{\alpha}\delta_{\beta,-\alpha} -
c^{\alpha +\beta}_{\alpha ,\beta}
\frac{ w_{\alpha}w_{\beta} }{ w_{\alpha +\beta} } A^{\alpha +\beta}_{k}
+(\alpha \cdot r^{\beta}) A^{\alpha}_{k} +
(\beta \cdot A^{\alpha}) A^{\beta}_{k}=0
\qquad
 (\forall k=1,\ldots, n),  
\label{p1}\ee
whereby we determine the constants $b_k^\alpha$ 
that appear in $A_k^\alpha =w_\alpha b_k^\alpha$ (\ref{formB}).
We here use the notation 
$\alpha \cdot r^{\beta}=\sum_{i=1}^{n}
\alpha_{i}r_i^{\beta}$, 
 $\beta \cdot A^{\alpha}=\sum_{i=1}^{n} \beta_{i}A^{\alpha}_{i}$
and similarly for all quantities with Cartan indices.
For later reference, note from (\ref{ria}) that 
\be
\beta \cdot r^{\alpha}={1\over 2} w_\alpha\beta\cdot  (C_\alpha - K_\alpha),
\qquad
\forall\alpha,\beta\in \Phi, 
\label{A*}\ee
where $K_\alpha$ is defined in (\ref{glnconv}) and
$C_\alpha = \sum_{i=1}^n C_\alpha^i H_i$ enjoys the properties  in (\ref{9}).

If we fix $\alpha\in \Phi$, then 
(\ref{p1}) for the pairs of roots $(\alpha,\beta)$ given by 
$(\alpha,\alpha)$, $(-\alpha,-\alpha)$, $(\alpha,-\alpha)$ and
$(-\alpha,\alpha)$
leads respectively to the following relations:
\be
(\alpha\cdot r^{\alpha} +\alpha\cdot A^{\alpha})A_k^{\alpha} =0,
\label{p2}\ee
\be
(\alpha\cdot r^{-\alpha} +\alpha\cdot A^{-\alpha})A_k^{-\alpha}=0,
\label{p3}\ee
\be
\alpha_k w^{2}_{\alpha} +(\alpha\cdot r^{-\alpha})A_k^{\alpha} -
(\alpha\cdot A^{\alpha})A_k^{-\alpha} =0,
\label{p4}\ee
\be
\alpha_k w^{2}_{\alpha} + (\alpha\cdot r^{\alpha})A_k^{-\alpha} - 
(\alpha\cdot A^{-\alpha})A_k^{\alpha} =0.
\label{p5}\ee
Since  
$\alpha\cdot r^\alpha = \alpha\cdot r^{-\alpha}$ 
by (\ref{A*}), these relations imply that 
\be
\alpha \cdot A^{\alpha}=\alpha\cdot A^{-\alpha} =-\alpha\cdot r^{\alpha}.
\label{p6}\ee
On account of (\ref{p6}) and (\ref{A*}), (\ref{p4}) can be written as
\be
\alpha_k w^{2}_{\alpha} =(\alpha\cdot A^{\alpha} )
(A_k^{\alpha} +A_k^{-\alpha}).
\label{p7}\ee
This expression shows that
\be
b_k^{\alpha} - b_k^{-\alpha} =\varepsilon^{\alpha}  \alpha_k
\label{p8}\ee
with some constants $\varepsilon^{\alpha}$. 
We then find from the above that
\be 
\alpha\cdot b^\alpha =\varepsilon^\alpha
\ee
and the $\varepsilon^\alpha$ must satisfy 
\be 
\varepsilon^{\alpha} =\varepsilon^{-\alpha},
\qquad
(\varepsilon^{\alpha})^2=1.
\label{p10}\ee
Now it is convenient to introduce  
$\Pi_k^\alpha:= (b_k^\alpha + b_k^{-\alpha})$, 
which results in 
\be
b_k^{\alpha}=
\frac{1}{2}\varepsilon^{\alpha}\alpha_k +\frac{1}{2}\Pi_k^{\alpha}, 
\qquad \forall\alpha\in\Phi.
\label{p15}\ee
Let us put  $\Pi^{ij}_{k}:=\Pi^{(\lambda_i -\lambda_j)}_{k} $.
Then the relations $\Pi_k^\alpha = \Pi_k^{-\alpha}$ and 
$\alpha\cdot \Pi^\alpha=0$ (by (\ref{p6})) 
give
\be
\Pi^{ij}_{k} =\Pi^{ji}_{k},\qquad \Pi^{ij}_{i} =\Pi^{ij}_{j},\qquad 
\forall k,\, i\neq j.
\label{p14}\ee

Consider now such roots 
$\alpha =(\lambda_i -\lambda_j )$ and 
$\beta =\pm (\lambda_l -\lambda_m ) \in \Phi$ that 
$\left\{ i,j\right\} \cap\left\{ l,m\right\} =\emptyset$.
In this case (\ref{p1}) yields 
\be
(\alpha\cdot \hat{r}^{\beta})b_k^{\alpha} +
(\beta\cdot b^{\alpha})b_k^{\beta}=0,
\label{p16}\ee
\be
(\alpha \cdot \hat{r}^{-\beta})b_k^{\alpha} -
(\beta\cdot b^{\alpha})b_k^{-\beta} =0,
\label{p17}\ee
where we use the notation  
$\hat{r}^{\gamma}:=
\frac{ r^{\gamma} }{ w_{\gamma} }$ for any $\gamma\in\Phi$. 
Adding these two equations, and 
using (\ref{p6}) and (\ref{p15}), 
we can easily get that now
\be
\beta\cdot b^{\alpha}=0,\qquad \beta\cdot \Pi^{\alpha}=0.
\label{p18}
\ee
The general form of $\Pi_k^{ij}$ which obeys (\ref{p14}) and
(\ref{p18}) is in fact the following:
\be
\Pi_k^{ij} =\eta^{\alpha} (\delta_{ki} +\delta_{kj}) +2\Omega^{\alpha},
\label{p*}\ee
where $\eta^{\alpha}$, $\Omega^{\alpha}$ are constants.
Notice that for $\alpha=(\lambda_i - \lambda_j)$  that element 
$K_{\alpha}=\sum_{k=1}^n  K_\alpha^k H_k$ 
defined in (\ref{glnconv}) has precisely the components 
$K_{\alpha}^k = \delta_{ki}+ \delta_{kj}$.  

Now, let $\alpha, \beta, \alpha +\beta \in\Phi$ be roots. 
In this case $\alpha-\beta =\alpha
+(-\beta)\notin\Phi$.
Hence (\ref{p1}) for the $(\alpha,\beta)$ and the $(\alpha,-\beta)$ 
pairs reads as 
\be
c^{\alpha +\beta}_{\alpha,\beta} b_k^{\alpha +\beta} =
(\alpha\cdot \hat{r}^{\beta})b_k^{\alpha} +(\beta\cdot b^{\alpha}
)b_k^{\beta},
\label{p20}\ee
\be
0=(\alpha \cdot \hat{r}^{-\beta})b_k^{\alpha} -
(\beta\cdot b^{\alpha})b_k^{-\beta}.
\label{p21}\ee
By adding these two equations making 
use of (\ref{A*}) and (\ref{p8}), we obtain
\be
c^{\alpha +\beta}_{\alpha, \beta} b_k^{\alpha +\beta} =- 
(\alpha\cdot K_\beta) b_k^{\alpha} +
\varepsilon^{\beta}  (\beta\cdot b^{\alpha}
) \beta_k.
\label{p22}\ee
If $\alpha=(\lambda_i -\lambda_j)$, $\beta=(\lambda_j -\lambda_l)$ are 
chosen, then 
$\alpha\cdot K_\beta=-1$ and $c^{\alpha+\beta}_{\alpha, \beta} =1$.
Let us then substitute (\ref{p15}) with (\ref{p*}) into (\ref{p22}) 
and consider the resulting equation for  
$k\notin\{i,j,l\}$ and for $k\in \{i,j,l\}$. 
In this way we obtain the 
requirements\footnote{We here implicitly assume that $n\geq 4$, 
but the final solution is valid for any $n\geq 2$.}
\be
\Omega^{\alpha +\beta} =\Omega^{\alpha},
\label{e1}\ee
\be
\varepsilon^{\alpha+\beta}+\eta^{\alpha +\beta}=
\varepsilon^{\alpha} +\eta^{\alpha},
\label{e2}\ee
\be
\varepsilon^{\alpha} -\eta^{\alpha} = 
2\varepsilon^{\beta} (\beta \cdot b^{\alpha}),
\label{e3}\ee
\be
\eta^{\alpha +\beta}-\varepsilon^{\alpha +\beta}  = 
-2 \varepsilon^{\beta} (\beta \cdot b^{\alpha}).
\label{e4}\ee
These tell us that 
\be
\Omega^{\alpha +\beta}=\Omega^{\alpha}, \quad
\varepsilon^{\alpha+\beta}=\varepsilon^{\alpha},
\quad
\eta^{\alpha +\beta}=\eta^{\alpha}.
\label{p24}\ee
In conclusion, there exist some constants  $\varepsilon$, $\eta$, $\Omega$ that
\be
\varepsilon^{\alpha}=\varepsilon, 
\quad
\eta^{\alpha}=\eta,\quad
 \Omega^{\alpha}=
\Omega,\qquad \forall\alpha\in\Phi.
\label{p25}\ee
In addition, we can compute from (\ref{p15}) that in the above case 
$2\beta \cdot b^\alpha=(\eta-\varepsilon)$, and  by
 substituting this back into (\ref{e3}) we obtain
\be
(\varepsilon +1)(\eta -\varepsilon)=0.
\label{p26}\ee
At the same time we know from (\ref{p10}) that $\varepsilon$ 
must be equal to $1$ or $-1$.

The first solution of (\ref{p26}) is $\varepsilon=1=\eta$.
In this case we can determine $b_k^\alpha$ from (\ref{p15})  
in terms of the arbitrary 
constant $\Omega$ as
\be
b_k^{\lambda_{i} -\lambda_{j}}=\delta_{ki}  +\Omega.
\ee 
We can then also calculate $\beta \cdot r^\alpha$ from the above equations, 
and thereby find from (\ref{A*}) that $C_\alpha = - H_\alpha$ must hold.
This is precisely the result stated in case I of Proposition 2.
We have obtained it as a consequence of considering a subset of all cases 
of (\ref{p1}), but it can checked to satisfy this equation in
all remaining cases 
(for $\alpha=(\lambda_i -\lambda_j)$, $\beta=(\lambda_l - \lambda_i)$ etc.)
as well.
 
The other solution of (\ref{p26}) is $\varepsilon=-1$, but 
then we still have to determine $\eta$.
For this we consider $\alpha=(\lambda_i -\lambda_j)$, 
$\beta=(\lambda_j-\lambda_l)$ and calculate  that 
\be
b_k^{\alpha}=\frac{1}{2}\left(\left(\eta -1\right) \delta_{ki} +
\left( \eta +1\right) \delta_{kj}\right) +\Omega,
\label{ve1}\ee
\be 
b_k^{\beta}=\frac{1}{2}\left( \left( \eta -1 \right) \delta_{kj} + 
\left( \eta +1\right) \delta_{kl} \right) 
+\Omega .
\label{ve2}\ee
We then look at (\ref{p1}) for the 
$(\alpha ,\beta )$ and $(\beta ,\alpha)$ pairs of roots and add 
these two equations,
which gives 
\be
0=(\alpha\cdot \hat{r}^{\beta} +\alpha \cdot b^{\beta})b_k^{\alpha} 
+(\beta\cdot \hat{r}^{\alpha} +\beta\cdot b^{\alpha})b_k^{\beta}.
\ee
Since $b_k^\alpha$ and $b_k^\beta$ are linearly independent 
$n$-component vectors for any $\eta$,
we obtain  
\be
\alpha\cdot \hat{r}^{\beta} +\alpha\cdot b^{\beta} =0, 
\qquad 
\beta \cdot \hat{r}^{\alpha} +\beta\cdot b^{\alpha} =0.
\ee
By subtracting these equations and taking into account that by 
(\ref{A*}) now 
\be
\alpha\cdot \hat r^\beta - \beta\cdot \hat r^\alpha = {1\over 2} 
\left(\beta \cdot K_\alpha - \alpha\cdot K_\beta\right) = 1,
\ee
we find that $\eta =1$.
So we have completely determined $b^{\alpha}_{k}$ again, and 
it is easy to confirm that the final formula agrees with case II of 
Proposition 2.
Thus the proof is complete.

\section{Proof of Proposition 5}
\setcounter{equation}{0}
\renewcommand{\theequation}{C.\arabic{equation}}

In this appendix we verify the statement of Proposition 5.

By combining eq.~(\ref{Rprime}) and Proposition 4, the constant r-matrix 
that we wish to calculate can be written in the form 
\be
\tilde r'= \left( \varphi (q) \otimes \varphi (q)\right) \rho(q)  
\left( \varphi (q) \otimes \varphi (q) \right)^{-1}
\ee
with 
\be
\rho(q)=\left( \chi(q) \otimes \chi(q)\right) \left(\tilde r(q) + 
\sum_{k} A_k(q) \otimes H_k \right) 
\left(\chi(q) \otimes \chi(q)\right)^{-1}.
\ee
The formulas in (\ref{RA}), (\ref{chi}) together with 
(\ref{lemma2}) and (\ref{lemma1}) result in      
\bea
&&\rho = -{\cal B}\sum_{k \neq l} \frac{1}{F_k -F_l }
(e_{kl} -e_{ll} )\otimes (e_{lk} -e_{kk} )
+\sum_{k \neq l}\frac{F_{k} F_{l} }{F_k -F_l}
(e_{kl} -e_{ll} ) \otimes (e_{lk} -e_{kk} ) 
\nonumber\\
&& \qquad  
+\sum_{k \neq l}F_k e_{kk} \otimes e_{kl} 
- \sum_{k \neq l} F_l e_{lk} \otimes e_{ll}. 
\label{rhoq}\eea
Therefore, to prove Proposition 5 it is enough to verify that 
\be
\left(\varphi (q) \otimes \varphi (q)\right) \rho(q) 
=\tilde r' \left(\varphi (q) \otimes \varphi (q)\right)
\label{wish}\ee
holds for $\rho$ in (\ref{rhoq}) and $\tilde r'$ in (\ref{constR}). 
We obtain in a straightforward manner that  
\be
\left(\varphi (q) \otimes \varphi (q)\right) \rho(q) 
=\sum_{a,b,c,d=1}^n 
\left( {\cal B}B_{abcd} +{\tilde B}_{abcd} \right) e_{ab} \otimes e_{cd},
\ee
where 
\be
B_{abcd}=
\frac{(\varphi_{ad} -\varphi_{ab} )(\varphi_{cd} -\varphi_{cb})}{F_d -F_b }, 
\quad\mbox{if}\quad b \neq d, 
\label{B}\ee
\be
\tilde{B}_{abcd}=
\frac{F_d F_b }{F_d -F_b }
(\varphi_{ad} -\varphi_{ab} )(\varphi_{cd} -\varphi_{cb})
+F_{d} \varphi_{ad} \varphi_{cd} -F_{b} \varphi_{ab} \varphi_{cb},
\quad\mbox{if}\quad b \neq d,
\ee
and $B_{abcd}={\tilde B}_{abcd}=0$ if $b=d$.
{}From (\ref{varphi}) and (\ref{constR}),
the right hand side of (\ref{wish}) is found to be
\be
 \tilde r' \left(\varphi (q) \otimes \varphi(q)\right) =\sum_{a,b,c,d=1}^n 
\left( {\cal B} D_{abcd} +\tilde{D}_{abcd} \right) e_{ab} \otimes e_{cd}
\ee
with
\be
D_{abcd}=\sum_{(a,x,c,y) \in S}\varphi_{xb} \varphi_{yd} -
\sum_{(c,y,a,x) \in S}\varphi_{xb} \varphi_{yd},
\label{D}\ee
\be
\tilde{D}_{abcd}=\sum_{(a-1,x,c-1,y) \in S}\varphi_{xb} \varphi_{yd} -
\sum_{(c-1,y,a-1,x) \in S}\varphi_{xb} \varphi_{yd},
\ee
where the set $S$ is defined in Proposition 5 and by an empty 
sum we mean zero.

We now observe that
$\tilde{D}_{abcd}=0=\tilde{B}_{abcd}$
if $a=1$ or  $c=1$, and
\be
\tilde{D}_{a,b,c,d}=D_{a-1, b, c-1, d},
\qquad
\tilde{B}_{a,b,c,d}=B_{a-1, b, c-1, d},
\qquad\mbox{if}\quad
2\leq a,c\leq n.
\label{BD}\ee 
These properties are obvious for $\tilde D$, 
while for $\tilde B$ they follow from the formula (\ref{varphi}).
In particular, the second equality in (\ref{BD}) 
is checked by inserting into (\ref{B}) the identity 
\be
\varphi_{a-1,d} -\varphi_{a-1,b} =F_{b}\varphi_{ab} -F_{d} \varphi_{ad},
\qquad
2\leq a \leq n,
\ee
which is consequence of (\ref{varphi}).
We conclude that it is sufficient to show 
that $B_{abcd} =D_{abcd}$.

Let us examine the expressions of $B_{abcd}$ and $D_{abcd}$.
First, we notice that for all indices 
\be
B_{abcd}=B_{cbad},
\qquad
D_{abcd}=D_{cbad},
\ee
and 
\be
B_{abcd}=0=D_{abcd} 
\quad\mbox{if}\quad a=n 
\quad\mbox{or}\quad c=n
\quad\mbox{or}\quad  b=d.
\ee
Hence it is enough to show that $B_{abcd}=D_{abcd}$ for 
such indices that $a\leq c<n$ and $b\neq d$.
We now introduce the notation 
\be
F_P:= \prod_{t\in P} F_t
\qquad
\forall P\subset \{1,\ldots, n\},
\ee
and also put $F_P:=1$ if $P=\emptyset$, for which $\vert P\vert=0$.
We then rewrite $B_{abcd}$ as
\be
B_{abcd}=
(F_{d} -F_{b} )
\Bigl(
\sum_{\begin{array}{c}
P\subset I^n_b \cap I^n_d \\
\vert P\vert=n-1-a
\end{array}}
F_P
\Bigr)
\Bigl(
\sum_{\begin{array}{c} P\subset I^n_b \cap I^n_d\\
\vert P\vert=n-1-c\end{array}} 
F_P 
\Bigr),
\label{Bexpand}\ee
where $I^n_k$ is defined in (\ref{I-set}). 
This is derived from (\ref{B}) by using that 
as a result of (\ref{varphi})
\be
\varphi_{al}-\varphi_{ak}= (F_k-F_l) 
\sum_{\begin{array}{c} P\subset I^n_k \cap I^n_l\\
\vert P\vert=n-1-a\end{array}} 
F_P.
\ee
Next, by inserting (\ref{varphi}) into (\ref{D}) 
and using that  $a\leq c$,  we get the expression 
\bea
&& D_{abcd}=(F_{d} -F_{b})
\sum_{\begin{array}{c} x+y=a+c+1\\
 1\leq x\leq a<y\leq n
\end{array}}
\Biggr(
\Bigl(
\sum_{\begin{array}{c} P\subset I^n_b\cap I^n_d \\
\vert P\vert=n-1-x
\end{array}}
F_P 
\Bigr)
\Bigl(
\sum_{\begin{array}{c} P\subset I^n_b\cap I^n_d\\
\vert P\vert=n-y\end{array}}
F_P 
\Bigr)
\nonumber\\
&&\qquad \qquad 
-\Bigl( \sum_{\begin{array}{c}P\subset I^n_b\cap I^n_d\\ 
\vert P\vert=n-1-y\end{array}}
F_P
\Bigr) 
\Bigl( \sum_{\begin{array}{c} P\subset I^n_b\cap I^n_d\\
\vert P\vert=n-x\end{array}}
F_P 
\Bigr)
\Biggr).
\label{Dexpand}\eea
The $x=a$, $y=(c+1)$ term in the first line 
of the right hand side of (\ref{Dexpand}) 
clearly equals the right hand side of (\ref{Bexpand}).
The proof is completed by a 
close inspection of the ranges of the summation indices,
which  shows that all the remaining terms cancel pairwise between 
the two lines of (\ref{Dexpand}) for any $a\leq c\leq (n-1)$.

\end{document}